\documentclass[prl,aps,twocolumn,epsfig]{revtex4}
\usepackage{graphicx}% Include figure files
\usepackage{dcolumn}% Align table columns on decimal point
\usepackage{bm}% bold math
\makeatletter

\newcommand{\Rmnum}[1]{\expandafter\@slowromancap\romannumeral #1@}
\makeatother
\begin{document}
\title {Suppression of low-frequency noise in two-dimensional electron gas at
degenerately doped Si:P $\delta-$layers}
\author{Saquib Shamim,$^1$$^,$\footnote[1]{email:saquib@physics.iisc.ernet.in} Suddhasatta Mahapatra,$^2$ Craig Polley,$^2$ Michelle Y. Simmons,$^2$
and
Arindam
Ghosh$^1$}
\vspace{1.5cm}
\address{$^1$ Department of Physics, Indian Institute of Science, Bangalore 560 012, India}
\address{$^2$ Center for Quantum Computer Technology, University of New South Wales, Sydney NSW 2052, Australia}
%\date{\today}
\begin{abstract}
We report low-frequency $1/f$ noise measurements of degenerately doped Si:P
$\delta$-layers at $4.2$~K. The noise was found to be over six orders of magnitude
lower than that of bulk Si:P systems in the metallic regime and is one of the lowest values reported for doped semiconductors. The noise was found to be nearly independent of magnetic field at low fields, indicating negligible contribution from universal conductance fluctuations. Instead interaction of electrons with very few active structural two-level systems may explain the observed noise magnitude.
\end{abstract}

%/PACS:  75.47.Gk,  73.40.-c,  75.70.-i

\maketitle

As classical information processing technology approaches the sub-20~nm node,
it is becoming increasingly important to control the exact number and position
of dopants in electronic devices~\cite{ITRS2009,Intel2008}. Recent progress in using scanning tunneling
microscopy (STM) as a lithographic tool allows positioning of dopants with
atomic scale precision~\cite{Schofield2003}. Combined with molecular beam epitaxy, this
technology has been employed to realize heavily $\delta$-doped, planar
nanostructures, such as tunnel gaps~\cite{Ruess2007A}, nanowires~\cite{Ruess2007} and quantum dots~\cite{Fuhrer2009,Fuechsle2010}. The same approach can also be used to fabricate vertically-stacked, multiple electrically-active layers~\cite{mckibbin2009}. The time averaged
transport properties of Si:P $\delta$-doped layers have now been studied in
detail~\cite{Goh2006,Ruess2007,Fuhrer2009,Fuechsle2010}, but very little is
known about its long term charge stability, which reflects in low frequency flicker noise in the electrical transport. The importance of this issue is paramount to the
overall development of devices with controlled dopant positioning at the nano-scale and in particular for single dopant spin based qubits~\cite{Hollenberg2004PRB}.

The noise properties of bulk doped Si has been studied in the metallic regime as well as near the metal-to-insulator transition (MIT). While universal conductance fluctuations (UCF) has been observed in the metallic samples~\cite{AGPRL2000}, signatures of glassy behavior has been seen near the MIT~\cite{SwastikPRL2003}. The situation is far more unclear when the dopants are confined wihtin one or very few atomic layers. This acquires additional significance due to stronger interaction effects at lower dimensions, theoretical predictions of exotic magnetic states~\cite{BhatPRB2007,BhatIJMPB2007} and other possibilities of Hubbard physics close to half filling that are naturally realized in these $\delta$-doped Si systems.

In this Letter we present the study of low-frequency noise, or $1/f$-noise, in
degenerately doped Si:P $\delta$-layers. We perform noise measurements as a function of number of carriers to establish that the measured conductivity fluctuations originate from the $\delta$-layer. We find that the noise is many orders of magnitude lower than bulk doped metallic silicon. Though the magneto-conductivity data indicates weak localization (WL), the magnetic field dependence of noise speaks against any significant contribution from UCF. Instead, interaction of electrons with a small concentration of tunneling two-level-systems (TLS) may explain the extremely low noise in these heavily doped systems.

\begin{figure}[t]
%\begin{center}
\includegraphics[width=1\linewidth]{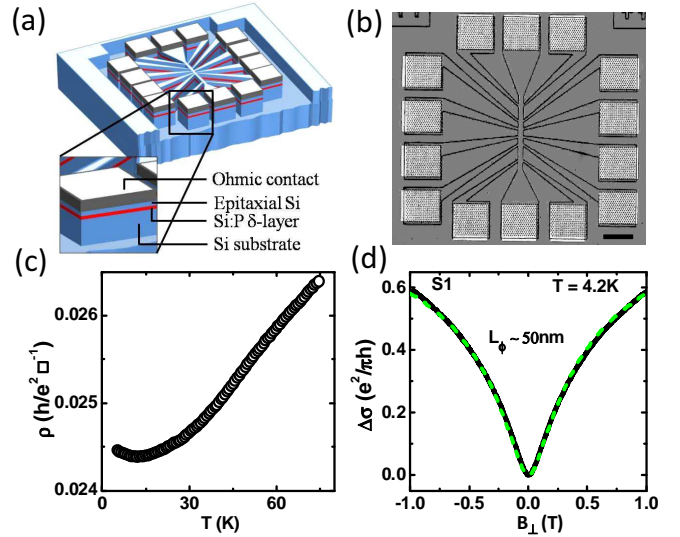}
\caption{(Color online)(a) Schematic of the device structure showing the $\delta$-layer(red line) of P atoms inside Si.  (b) Optical image (false color) of device S2 used in this experiment. The scale bar is $100\mu$m (c) Resistivity vs. temperature, $T$ for a device similar to the one studied in this work. (d) Magneto-conductivity plot for sample S1 at T = $4.2$~K. The dashed line is the weak localization fit to the data.}
%\end{center}
\end{figure}

Two Hall bars (S1 and S2) from the same $\delta$-doped Si:P layer have been studied in
this work. The Si:P $\delta$-layer was fabricated in an ultra-high vacuum variable-temperature STM system equipped with a phosphine (PH$_3$) dosing system and a Si sublimation source. The details of sample fabrication have been
reported elsewhere~\cite{Goh2006}. The trench-isolated Hall bars, were fabricated by electron-beam
lithography and reactive ion etching. Figure~1a shows a schematic of the final device structure, wherein a $\delta$-layer of P atoms is indicated by the red line. Both Hall bars have a width of 20 $\mu$m and multiple voltage probes for carrier-number dependent noise measurements. Ohmic contacts to the Hall bars were made by depositing 60 nm of nickel (Ni) and 10 nm of titanium (Ti), followed by annealing in nitrogen atmosphere at 350~$^\circ$C and depositing another layer of Ti/gold (Au) (10nm and 60nm respectively). An optical image of the device S2, recorded after deposition of the ohmic contacts, is shown in Figure 1b.

\begin{figure}[t]
%\begin{center}
\includegraphics[width=1\linewidth]{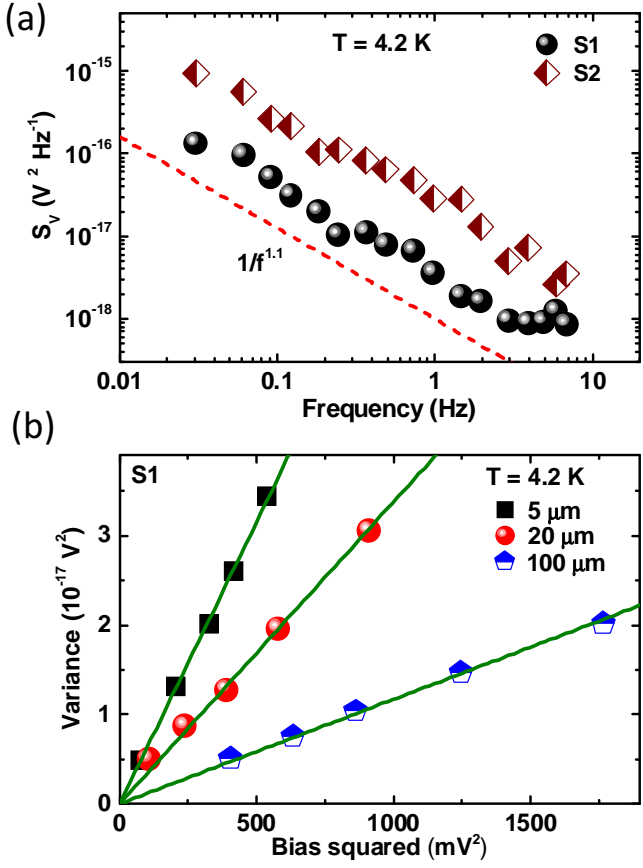}
\caption{(Color online)(a) The power spectral density (PSD), $S_V$ as a function of frequency for both devices at T = $4.2$~K and zero magnetic field ($B_\bot$ = 0). The dashed line indicates that the spectrum is ~$1/f$ in nature. The spectrum of S2 is offset by three times for visual clarity. (b) The PSD, $S_V$ as a function of $V^2$ for three different channel lengths of sample S1. The solid lines shows linear fits to the data. }
%\end{center}
\end{figure}

From Hall measurements at temperature $T = 4.2$~K, the two dimensional (2D) electron
density of both samples was estimated to be $n = (1.22\pm0.01)\times10^{14}$~cm$^{-2}$. Both devices indicate a
finite residual resistivity $\rho_0 \sim 600 ~\Omega$/square, with $k_F\ell \simeq 40$ ($k_F$ and $\ell$ are Fermi
wavevector and mean scattering lengths, respectively). The initial decrease in
resistivity $\rho$ with decreasing $T$ (from $T = 70$~K to 12~K) in both
devices (data for S2 not shown) confirms metallic-like behavior(Figure 1c). The upturn in $\rho$ for $T \lesssim
12$~K is associated with the WL effect. To probe this
further, and also extract the phase relaxation length $L_\phi$, we have
performed four probe magneto-conductance measurement at $4.2$~K. Figure~1d shows the magneto-conductivity
plot of S1, measured in a perpendicular magnetic field, ($B_\perp$) at 4.2K. The dip at
$B_\perp = 0$ is the hallmark of WL behavior. The magneto-conductivity data was fitted with the Hikami formulation~\cite{Hikami} for disordered 2D systems, which gives a phase breaking field ($B_\phi$) of $\sim $60mT and $L_\phi\sim$~50nm at $4.2$~K. For both resistance and noise measurements, the voltage drop across the device, $V$ was kept $\ll(k_BT/e)L/L_\phi$ to minimize heating of electrons.

\begin{figure}[b]
%\begin{center}
\includegraphics[width=1\linewidth]{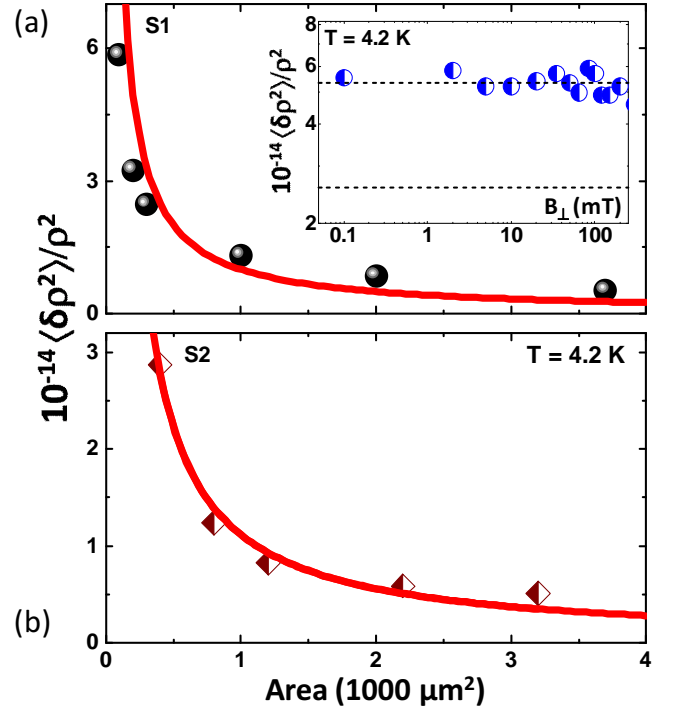}
\caption{(Color online)(a) The normalized variance $\langle$$\delta$$\rho^2$$\rangle$/$\rho^2$ as a function of area for S1 at T = $4.2$~K and $B_\bot$ = 0. (Inset) $\langle$$\delta$$\rho^2$$\rangle$/$\rho^2$ vs $B$ at T = $4.2$~K for S1. The dashed lines show average value for $\langle$$\delta$$\rho^2$$\rangle$/$\rho^2$ and $1/2\times\langle$$\delta$$\rho^2$$\rangle$/$\rho^2$. (b) $\langle$$\delta$$\rho^2$$\rangle$/$\rho^2$ as a function of area for S2 at T = $4.2$~K and $B_\bot$ = 0. The solid lines represent $1/A$ variation of noise magnitude. }
%\end{center}
\end{figure}

For noise measurements we used an AC four-probe wheatstone bridge
technique~\cite{Scofield1987,AGarxiv,AGPRL2000}. The voltage drop across the sample was amplified by a low noise voltage preamplifier (SR 560) and the output of the amplifier was balanced across a standard wire wound resistor. The voltage fluctuations
were recorded as a function of time using a 16 bit digitizer. The raw data was then processed digitally using a three stage decimation process, followed by the
power spectral density (PSD) estimation. The details of the noise measurement process
can be found elsewhere~\cite{Scofield1987,AGarxiv}. The PSD, $S_V$($f$), of noise as a function
of frequency $f$ is shown in Figure~2a for both the samples. In both devices, we
found $S_V \propto 1/f^\alpha$, where the frequency exponent $\alpha \approx
1 - 1.2$ over the entire experimental bandwidth. The bias dependence of $S_V$ shown in Figure~2b was recorded for different distances between the voltage probes for sample S1. The solid lines show linear fits to the data. $S_V$ was found to
be $\propto V^2$, for all cases, which ensures that we are in the ohmic regime, where the
measured voltage fluctuations represent the fluctuations in $\rho$ of the
Si:P $\delta$-layer {\it i.e.} $S_V/V^2 = S_\rho/\rho^2$. Moreover, the slope of the linear fits decreases as the separation between the voltage probe increases, which essentially means that $S_\rho$ decreases as the total number of carriers increases.

The frequency and the bias dependence of noise can be combined to normalize the noise magnitude in terms of the
phenomenological Hooge relation,

\begin{equation}
\label{eq1} \frac{S_\rho(f)}{\rho^2} = \frac{\gamma_H}{nAf^\alpha}
\end{equation}

\noindent where $\gamma_H$, $n$ and $A$ are the phenomenological Hooge
parameter, areal density of electrons, and
the area of the Hall bar between the voltage probes, respectively. In the data
shown in Figure~2, the magnitude of $\gamma_H$ was deduced to be around
$10^{-6}$, which is orders of magnitude lower than that of bulk doped
Si:P systems degenerately doped to the metallic regime ($\gamma_H$ = 0.1-2)~\cite{AG99,Swastik2001}. Given such a low value of $\gamma_H$, it is important to establish that we indeed are measuring the noise from the $\delta$-layer. Noise measurements were performed for different distances between the voltage probes for both S1 and S2. The results are shown in Figures~3a and 3b, respectively, where $\langle$$\delta$$\rho^2$$\rangle$/$\rho^2$ is plotted against the area ($A$) of the $\delta$-layer between the voltage probes (here, $\delta\rho^2 = \int S_\rho(f)df$). As expected from the Hooge relation $\langle$$\delta$$\rho^2$$\rangle$/$\rho^2$ shows a 1/$A$ dependence, confirming that the measured resistance fluctuations come from the Si:P $\delta$ layer, where different fluctuators contribute independently to the observed noise magnitude.

In order to understand the microscopic origin of noise in Si:P $\delta$-layers we have investigated the noise magnitude as a function of perpendicular magnetic field ($B_\perp$) for S1 at T = 4.2K. We find that the $\gamma_H$ remains essentially constant over the range corresponding to the phase breaking field $B_\phi$(Inset of Fig.~3a) which is $\sim$ 60 mT at T = 4.2K. The near constancy of noise with $B_\perp$ shows that universal conductance fluctuations (UCF) as a major source of noise is quite unlikely as we do not see any factor of $2$ reduction in $\gamma_H$~\cite{BirgePRB1993,AGPRL2000} expected on removal of time reversal symmetry. This is a surprising result since magnetoresistance clearly displays WL (Fig.~1d) and UCF and WL are expected to be manifestations of same quantum interference effect. It is however possible that the low temperature Hamiltonian of our $\delta$-doped Si system has very different symmetry properties from conventional disordered conductors, although there is no clear understanding of why this should be so.

An alternative possibility is based on a local interference model suggested by
Kogan and Nagaev~\cite{Kogan1984} with anisotropic scattering of electrons by tunneling TLS. In
our system the TLS may be  associated with the incorporation of P-atoms in the
silicon matrix, as indicated in the bulk-doped systems~\cite{AGPRL2000}. In
this model, the resistivity fluctuations can be expressed as~\cite{Kogan1984}

\begin{equation}
\label{eq2} \frac{S_\rho(f)}{\rho^2} \approx \frac{n_{TLS}(\ell \sigma_s)^2}{Af}
\end{equation}

\noindent where $n_{TLS}$ and $\sigma_s$ are the areal density of the TLS and
scattering cross section of the electrons, respectively. Note that $\gamma_H$ in Equation (2) is independent of B. Taking $\sigma_s \sim
1/\sqrt{n_P}$, we estimate $n_{TLS}/n_P \simeq 2\pi\gamma_H(\rho_0/(h/e^2))^2
\sim 3\times10^{-7}$. This indicates that the magnitude of the observed noise can be explained if only $3$ in $10$ million P atoms form active TLS within the experimental bandwidth. Note that charge fluctuations on even a small number of defects can have a $1/f$ spectrum as long as relaxation rates are widely distributed~\cite{Pellegrini20001775}.

\begin{figure}[t]
%\begin{center}
\includegraphics[width=1\linewidth]{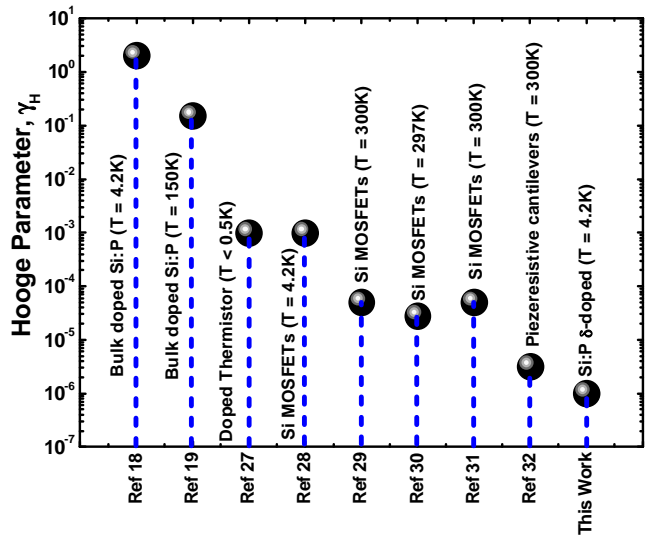}
\caption{(Color online) Comparison of various reported values of Hooge parameter,  $\gamma_H$ for doped silicon. Since the noise magnitude can strongly depend on temperature (T), the values of T for each reference have been explicitly stated.}
%\end{center}
\end{figure}

It is well known that at extremely high dopant concentrations defects can occur creating deactivating centres either through the formation of  donor vacancy clusters, donor pairs or donor pair vacancy interstitials ~\cite{DonorComplexes2003,ChadiPRL97,MuellerPRB2004,MoonNanolett2008}. However using a gaseous dopant source with self limiting absorption of the gaseous dopant precursor it is somewhat surprising that such defect complexes would occur. Other possible explanations could be the incomplete incorporation of dopants into substitutional sites during the incorporation anneal or the potential presence of defects in the epitaxial silicon overgrowth. These results highlight that despite the extremely low noise observed further work is needed to pinpoint exactly what gives rise to the noise floor in these heavily doped devices.

In Fig.~4 we compare the Hooge parameter of our system with the values reported previously for doped Si. Ghosh et. al.~\cite{AG99} and Kar et. al.~\cite{Swastik2001} have previously measured highly doped bulk Si:P systems and found fairly large values of $\gamma_H$ (0.1 to 2). In comparison to these bulk doped Si:P systems we find that noise in Si:P $\delta$-layers studied in this work are suppressed by 5 to 6 orders of magnitude. Other references included in  Fig.~4 are for doped thermistors~\cite{mccammon91}, Si MOSFETs~\cite{Adkins1982,gaubert394,vonHaartman2007771,Marin20041077} and piezoresistive cantilevers~\cite{yu6296}. An explanation of extreme low noise in our system may involve large elastic energy barriers around the P-atom which immobilizes them and reduces the number of active TLSs. Such barriers may arise during the doping process when the Si-Si bonds distort locally to incorporate the dopants. The remarkably low value of $\gamma_H$ measured here favors the use of Si:P $\delta$-doped devices as versatile nanoelectronic elements.

In conclusion, we demonstrated suppression of resistance noise in Si:P $\delta$-layers by several orders of magnitude, in comparison to degenerately doped bulk Si:P systems. The noise is nearly unaffected by low magnetic fields. We indicate the possible role of tunneling two-level systems within a local interference model to understand the microscopic origin of the noise.

We thank Department of Science and Technology (DST), Government of India and Australian-Indian Strategic Research Fund (AISRF) for funding the project. SS thanks CSIR for financial support.

\newpage
\noindent

\end{document}